\documentclass{article}

\usepackage{microtype}
\usepackage{amsfonts}
\usepackage{amssymb}
\usepackage{amsmath}
\usepackage{graphicx}
\usepackage{caption}
\usepackage{subfigure}
\usepackage{booktabs} 

\usepackage{hyperref}

\usepackage[accepted]{icml2019}

\icmltitlerunning{Decode and Transfer}

\begin{document}

\twocolumn[
\icmltitle{Decode and Transfer: A New Steganalysis Technique via Conditional Generative Adversarial Networks}



\icmlsetsymbol{equal}

\begin{icmlauthorlist}
\icmlauthor{Parisa Babaheidarian}{to}
\icmlauthor{Mark Wallace}{to}\\
Qualcomm Technologies Incorporation
\end{icmlauthorlist}

\icmlaffiliation{to}{Qualcomm Technologies Incorporation, USA}
\icmlcorrespondingauthor{Parisa Babaheidarian}{pbabahei@qti.qualcomm.com}

\icmlkeywords{Steganography, domain adaptation, GAN, image decoding, conditional generative adversarial networks, communication security}

\vskip 0.3in
]




\begin{abstract}
Recent work \citep{Baluja} showed that using a pair of deep encoders and decoders, embedding a full\- size secret image into a container image of the same size is achieved. This method distributes the information of the secret image across all color channels of the cover image, thereby, it is difficult to discover the secret image using conventional methods. In this paper, we propose a new steganalysis technique which achieves complete recovery of the embedded secret in steganography images. We incorporate a deep neural network to decode an approximate estimate of the secret image followed by a domain adaptation technique based on generative adversarial networks which transfers the decoded image into a high quality RGB image with details visible to human eyes. Our steganalysis technique can be served as an attack model against which the security level of an arbitrary embedded\- based digital watermarking or a steganography algorithm can be evaluated. Furthermore, our method can be used as a general framework to decode a high quality image message from a noisy observation of an encoded message.
\end{abstract}

\section{Introduction}
Steganography is an art and science of embedding a hidden message into a carrier signal. Unlike encrypted messages, a steganography message appears to be a plain message to a casual observer; it provides security through obscurity. It is possible to combine cryptography with steganography, e.g., by using a random secret permutation matrix as a key. There are two basic components in steganography, these are the secret message and the cover medium or container. The messages could be text, audio, images or anything that can be embedded in a bit stream. Images are usually the candidate for the container signal due to their redundancy which allows embedding secret messages more robustly and less perceptible. Most steganography techniques lead to a distortion in the container signal which can be used to detect the existence of an embedded secret. Mielikainen et al. proposed an improved version of least significant bit (LSB) matching that enables embedding the same payload as LSB matching but with fewer changes to the cover image \cite{MilkLSB}. Pevny et al. \cite{penvy} introduced highly undetectable stego (HUGO), a new embedding algorithm for spatial domain digital images based on minimizing a distortion function in some feature space.

Steganalysis is the process of discovering the hidden information. The rapid growth of statistical analysis and machine learning techniques has empowered steganalysis mechanisms which are the biggest threat to embedded based steganography algorithms. In \cite{Frid} Fridrich et al. proposed a reliable steganalysis method for detecting least significant bit (LSB) nonsequential embedding in digital images in which the length of the secret message length was derived by analyzing the lossless capacity in the LSB and
the shifted LSB plane. In \cite{penvy2} Penvy et al. developed a feature set based on discrete cosine transform (DCT) coefficients and Markov feature set to train a multi-label classifier which determines if there is an embedded secret in the container image for a few steganography algorithms. In \cite{goljan} Goljan et al. proposed an extension of the spatial rich model for steganalysis of color
images. The extended features are designed to capture dependencies across color channels.

Recently, deep steganalysis has been proposed which enhances statistical analysis of the container images. In \cite{qian}, Qian et al., proposed a customized convolutional neural network (CNN) for steganalysis to automatically extract features that are useful for steganalysis as opposed to complex handcrafted features. In \cite{zeng} a new hybrid steganalysis was proposed using a CNN framwork in which the CNN focuses on regions with complex textures. The motivation is to attack adaptive steganography algorithms which embed different number of bits in different regions of the image depending the local texture and complexity. To overcome some of these attacks, deep steganography techniques have been introduced recently \cite{dcgan} \cite{Baluja}. In \cite{dcgan} a new deep steganography without embedding was introduced in which the secret information is mapped into a noise vector and a generator neural network model is trained to generate the carrier image based on the noise vector. 

Most of the deep proposed methods attempted to convey a small size secret message through a carrier signal. However, in \cite{Baluja}, Baluja proposed a deep steganagraphy algorithm which embeds a full-size secret image into a carrier image of the same size. A pair of encoder and decoder deep neural networks are simultaneously trained to create the hiding and revealing processes. Baluja showed this method compresses and distributes the secret image's representation across all of the available bits unlike to commonly used LSB approach.

In this work, we analyze the steganography algorithm proposed in \cite{Baluja} and develop a new steganalysis framework which not only determines the existence of the embedded secret but also fully recovers the secret image with high peak signal to noise ratio and similarity index. Our steganalysis framework uses a deep CNN based decoder to decode the secret image given the original cover and container images, however, the decoded image is of poor quality and many of its details are invisible to human eyes. To remedy this, we transfer the poor quality decoded images into high quality RGB images via a domain adaptation technique known as pixel domain adaptation which was introduced in \cite{pixelda}. This domain adaptation technique transfers images from a source domain to images from a target domain using generative adversarial networks. We adapt the pixel domain adaptation model by adding additional cost terms to condition the generated images such that it suits our steganalysis problem. We show that our conditional generative adversarial model can learn the prior information about the domain of secret images which is effectively used to decode the embedded secret. Our simulations show that using limited examples of the steganography images, we can effectively train our steganalysis networks with fairly good generalization performance to unseen data. Also, since we do not make any assumption on the underlying stenography encoder, our framework can be generalized to analyze any embedded based deep steganography algorithm that preserves the spatial information of the secret image and does not incorporate any cryptographer key. As a result, our steganalysis framework can be used as an attack model against which the security level of an embedded based steganography/digital watermarking technique can be evaluated.

Moreover, with some modifications, our algorithm can be treated as an image decoder algorithm which successfully decodes images from a noisy channel with only few measurements. This ultimately impacts the compression and the transmission rates. In particular, using our decoder, the same quality of image decoding can be achieved with lower transmission rate than the data rate needed to successfully decode the image with a conventional decoder. 
\section{Background and Related Work}
In this section, we briefly review the deep steganography introduced in \cite{Baluja} and the pixel domain adaptation technique proposed in \cite{pixelda}.
\subsection{Deep Steganography}
\cite{Baluja} proposed a deep steganography method which used three deep neural networks for encoding and decoding the embedded secret image. Unlike previous methods, this method achieves hiding an $N \times N$ RGB secret image into an $N \times N$ RGB cover image with minimal visual distortion in the container image. A container image is a cover image in which the information of the secret image is hidden. Also, the discovery process is supposed to be with minimal distortion, but it is not designed to be lossless. The novelty of this method is that it essentially hides 10 to 40 times higher bit rates than the commonly used steganography algorithms. They showed that the information of the secret image is distributed across all color channels of the container image and it is not only in LSB positions. Therefore, it produces container images that are difficult to be visually detected. However, as the author suggested the existence of an embedded secret image could be determined using statistical analysis. 

\begin{figure*}[ht]
	\centering
		\includegraphics[width=0.6\textwidth]{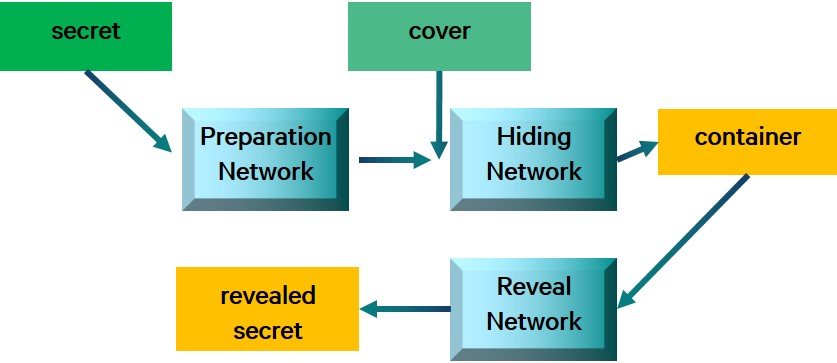}
	\caption{Illustration of the embedding and revealing procedures in the deep steganography algorithm \cite{Baluja} } 
	\label{figure:11}
\end{figure*}

The key idea in Baluja's algorithm is to train the encoder networks and the decoder network simultaneously. Baluja proposed a steganography model with three components: the Prep Network, the Hiding Network, and the Revealing Network. Figure \ref{figure:11} displays the procedures in this model. The Prep Network serves two main purposes: it resizes the secret image to the size of the cover image if they differ in size, and more importantly, it transforms the color-based pixels to more useful features which are later to be used in the embedding process. The Hiding Network takes the output of the Prep Network and the cover image as inputs and creates the container image. The goal is to create the container image such that the difference between the container and the cover images is visually imperceptible. Finally, the Revealing Network which is used by the legitimate decoder, receives only the container image and removes the cover image to reveal the embedded secret image. Given a set of cover and secret images, all three networks are trained simultaneously.
One possible scenario to analyze a steganography scheme is to assume the original cover image is accessed by the attacker. \cite{Baluja} showed that even though the difference between the original cover and container images can reveal limited information about the embedded secret image, revealing a visually meaningful copy of the original secret image is difficult even with 20 times image enhancement tools \cite{Baluja}.
\subsection{Pixel Domain Adaptation}
In \cite{pixelda} A team of researchers in Google Brain proposed a new domain adaptation technique using generative adversarial networks. The proposed technique learns a transformation in the pixel space from one domain to another. The input domain is called the source domain and the destination domain is called the target domain. In some sense, this technique is similar to the style transfer \cite{styletransfer} technique. However, following the style transfer approach, a style of a single image is learned which is then used to transfer the style of all the input images to the learned style whereas the pixel domain adaptation technique learns the style of an entire target domain and adapts the source domain images to appear as if they were drawn from the target domain. In \cite{pixelda}, the assumption is that the differences between the source and the target domains are primarily in low-level features which could include variation in noise level, texture, resolution, illumination, and color, as opposed to the high-level features such as types of objects, geometric variations, etc.
\section{The Proposed Steganalysis Framework}
Our steganalysis model follows the oracle attack model \cite{oracle} in which we assume given a pair of images (secret and cover images) the oracle outputs the container image according to the deep steganography  algorithm in \cite{Baluja}. Our assumption is the number of interactions with the oracle is limited. This means we have access to a limited size steganography training dataset to train our model.  Our goal is to find the inverse function which recovers the hidden image given original cover and associated container images. The attack is successful if not only the presence of an embedded secret is detected, but also, the secret image is recovered with high peak signal to noise ratio (PSNR) and high similarity to the original secret image, for any arbitrary test container image that has been produced by the same Steganography algorithm. It is worth to mention that while we demonstrate our attack against the deep steganography algorithm in \cite{Baluja}, our steganalysis algorithm can be generalized to attack any embedded-based steganography algorithm that preserves the spatial information of the embedded image and does not incorporate a cryptography key.

The key assumption in the steganography algorithm in \cite{Baluja} is that all three networks (Prep, Hiding and the Reveal Networks) are trained simultaneously which serves as the \textit{common randomness} among encoders and the decoder. In \cite{Baluja}, the encoders are basically the Preparation and Hiding Networks and the decoder is the Reveal Network. This common randomness allows future communications to be fully perceived by the Reveal Network but remain secret from an adversary's decoder. In our steganalysis model, we do not make any assumption on the encoders used in producing the steganography images which allows generalization of our attack to an arbitrary steganography algorithm with similar assumptions. However, we assume we have access to some side information regarding the target steganography algorithm. Specifically, we assume we have access to a limited dataset which contains tuples of (secret, cover, container) images where the container images were produced by the target steganography algorithm. These images are used to train our decoder.

Clearly, the absence of the common randomness in training the decoder leads to losing some information related to the embedded secret. Therefore, the decoded images are usually of poor quality and features low dynamic range which make it difficult for a human eye to perceive the details in the image. In image processing, this situation is known as an underdetermined problem. An underdetermined system is a system of equations that relate input to output for which the number of measurements (observations) is less than the number of unknowns. In our case, training the network using the limited number of steganography images lead to the same situation. A common approach to recover the signal from a set of underdetermined measurements is to incorporate some prior knowledge. This prior knowledge often assumes a prior distribution over the feasible solutions. For instance, a prior Gaussian distribution results in $\mathcal{L}_2$ regularization and Laplace distribution imposes $\mathcal{L}_1$ regularization. Other examples with applications to images are Tikhonov regularization \cite{tikh}, Lasso \cite{lasso}, and dictionary learning methods \cite{ksvd}.

Regularization techniques in training deep neural networks attempt to reduce overfitting to the training data by directly assuming a prior distributions on the network weights. However, in our steganalysis problem, we are interested in solutions certain characteristics, e.g. RGB images with high dynamic range, etc. Therefore, we assume a prior distribution over the estimated images produced by the network which indirectly implies a prior distribution over the networks' weights. Specifically, we propose the use of a domain adaptation technique based on generative adversarial networks (GANs) to learn the prior information regarding the secret images. Although GANs do not directly aim to learn the distribution, they are trained to produce samples that look like images drawn from the true distribution. GANs are known for generating sharp images which make them a suitable candidate for image enhancing tools. Motivated by the pixel domain adaptation technique in \cite{pixelda}, we use the GAN networks to transfer the poor quality decoded images to the domain of high resolution recovered images with high dynamic range which enables perceiving details that otherwise were not visible to human eyes even after applying common image enhancing tools \cite{Baluja}. We train the GAN networks by conditioning the solution of the generators to be similar to the original secret image in the least square sense. Our conditional GAN model ensures that the GAN learns low level features such as colors, textures, and resolution from the domain of possible secret images while preserving the spatial information of the embedded secret. This technique is similar to the pixel domain adaption technique introduced in \cite{pixelda} whose goal was to transfer images from one domain (e.g., binary MNIST domain) to another domain (e.g. colored MNIST-M domain) while preserving the foreground information.

Our discriminator network is trained on both generated images and random images sampled from the domain of high quality RGB images of the same size which share similar characteristic as the space of secret images, e.g., space of natural images. By balancing the coefficients of GAN's min-max cost function and the least square terms, we ensure that the generator learns how to transfer the poor quality decoded image to a high quality colored image with a rich dynamic range of colors and high similarity to the original secret.
\subsection{Model}
\begin{figure*}[ht]
	\centering
		\includegraphics[width=0.85\textwidth]{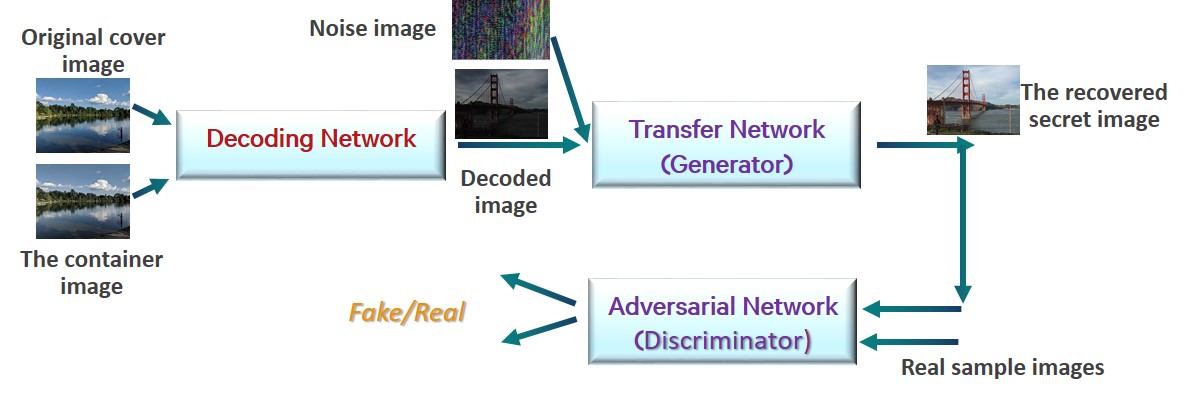}
	\caption{Our proposed Decode and Transfer steganalysis model.} 
	\label{figure:12}
\end{figure*}

Our steganalysis network architecture is illustrated in Figure \ref{figure:12}. The first part is our Decoding Network which takes the original cover and container images and attempts to decode an estimate of the original secret images by simply processing the differences between the two input image. We use a 5 layer convolutional neural networks (CNN) with ReLU activation functions for the Decoding Network. In our training dataset, we used GIF images, therefore, input and output images have 4 channels. The second part is our generative adversarial networks which we interchangeably refer to them as the Transfer and the Adversarial Networks, respectively. It takes the output of the Decoding Network and a random noise vector as inputs and outputs an estimate of the secret image. The input and output images of the Transfer Networks are similar in high level features (such as objects in the picture), however, they differ in image quality, noise level, colors, and textures. We adopt and customize the domain adaptation technique in \cite{pixelda} for modeling our Transfer (generator) and Adversarial Networks. The distinction is that we do not use a task classifier Network and the GAN's cost function is customized to serve our steganalysis purpose. Also, similar to the Decoding Network, the Transfer and Adversarial Networks work with images of 4 channels.
\subsection{Training}
Let $X^s, X^c, X^{c^{\prime}}, X^d$ and $X^{t}$ represent the random variables denoting secret, cover, container, decoded, and transferred image variables, respectively. Additionally, let $\mathbf{z}$ represent the random noise vector drawn from an i.i.d. Gaussian distributions. Also, assume $x^s, x^c, x^{c^{\prime}}, x^d, x^{t}$, and $z$ represent realizations of the aforementioned random quantities. Let $\theta_D, \theta_G$ and $\theta_A$ represent the trainable parameters of the Decoding, the Generative (transfer), and the Adversarial (discriminator) Networks, respectively.

Assume $D_{\theta_D}: X^c \times X^{c^{\prime}}\rightarrow X^d$ denotes the parameterized function representing the Decoding Network which. Also, assume $G_{\theta_G}: X^d \times \mathbf{z}\rightarrow X^t$ is the parameterized generator function that takes the decoded image and the noise vector as inputs and outputs a transferred image which is our final recovered secret image. Assume $X^r$ is the random variable denoting real images with distribution $P_{x^r}(x^r)$. A secret image $x^s$ is sampled from $P_{x^r}(x^r)$. Then, the Adversarial network is represented by the parameterized function $A_{\theta_A}: X^r \cup X^{t} \rightarrow \{0,1\}$, which decides whether its input image is real. A fake image is an image produced by the generator.

Let us define $\mathcal{L}_d(D_{\theta_D})$ as the decoding loss, $\mathcal{L}_t(A_{\theta_A},G_{\theta_G})$ as the domain transfer loss, and finally $\mathcal{L}_c(G_{\theta_G})$ as the conditional loss imposed on the images produced by the generator. Also, assume $\alpha, \beta$, and $\gamma$ are trade-off parameters in the overall cost function. Our goal is to optimize the following total loss functions:
\begin{equation}
\min_{\theta_D, \theta_G} \max_{\theta_A} ~\alpha \mathcal{L}_d(D_{\theta_D})+\beta\mathcal{L}_t(A_{\theta_A},G_{\theta_G})+\gamma \mathcal{L}_c(G_{\theta_G})
\end{equation}

The decoding loss function penalizes the distance between the output of the decoder network $x^d$ and the corresponding secret image $x^s$ in the least square sense. The decoding loss ensures the high level features in the input image to the generator network resemble the ones in the original secret image. We have
\begin{equation*}
\mathcal{L}_d(D_{\theta_D})\triangleq \mathbb{E}_{x^d, x^s} \bigg [\big \|x^d-x^s \big \|_2^2\big |\theta_D\bigg]
\end{equation*}
\begin{equation}
=\mathbb{E}_{x^c, x^{c^{\prime}},x^s}\bigg [\big \|D_{\theta_D}\left(x^c,x^{c^{\prime}}\right)-x^s \big \|_2^2 \big | \theta_D\bigg ]
\end{equation}

The domain transfer loss function is the GAN's min-max loss as it was formulated in \cite{pixelda}. We have
\begin{equation*}
\mathcal{L}_t(A_{\theta_A},G_{\theta_G})\triangleq \mathbb{E}_{x^s}\bigg [\log(A_{\theta_A}(x^s)\big |\theta_A \bigg ]+
\end{equation*}
\begin{equation*}
\mathbb{E}_{x^d, z}\bigg [\log (1-A_{\theta_A}(G_{\theta_G}(x^d,z)))\big | \theta_A, \theta_G\bigg ],
\end{equation*}
which is equivalent to
\begin{equation*}
\mathcal{L}_t(A_{\theta_A},G_{\theta_G})\triangleq \mathbb{E}_{x^s}\bigg [\log(A_{\theta_A}(x^s)\big |\theta_A \bigg ]+
\end{equation*}
\begin{equation}
\small
\mathbb{E}_{x^c, x^{c^{\prime}},x^s, z}\bigg [\log \left(1-A_{\theta_A}(G_{\theta_G}(D_{\theta_D}(x^c,x^{c^{\prime}}),z))\right)\bigg | \theta_A, \theta_G\bigg ]
\end{equation}
Lastly, the conditional loss function penalizes the mismatch between the transferred (generated) image with the original secret image. To further smooth the generated images, we add the Total Variation (TV) penalty to the overall cost function which further stabilizes the optimization over the domain transfer loss and gravitates the final solutions to less noisy images. Therefore, the conditional loss function is formulated as
\begin{equation*}
\small
\mathcal{L}_c(G_{\theta_G})\triangleq \mathbb{E}_{x^d, x^s, x^t} \bigg [\big \|x^t-x^s \big \|_2^2+ \lambda \big \|\nabla x^t\big \|^2_2 \bigg | \theta_D, \theta_G\bigg ]
\end{equation*}
\begin{equation}
\small
=\mathbb{E}_{x^d, x^s, z} \bigg [\big \|G_{\theta_G}(x^d,z)-x^s \big \|_2^2+ \lambda  \big \|\nabla G_{\theta_G}(x^d,z)\big \|^2_2 \bigg | \theta_D, \theta_G \bigg]
\end{equation}

which can further be expanded in terms of input quantities $x^c$ and $x^{c^{\prime}}$ and function $D_{\theta_D}(.)$. The gradient over $x^t$ in the TV cost is the two-dimensional differentiation operator on the image space which penalizes the horizontal and vertical image gradients and encourages smoother solutions.

\begin{figure}
	\centering
		\includegraphics[width=0.44\textwidth]{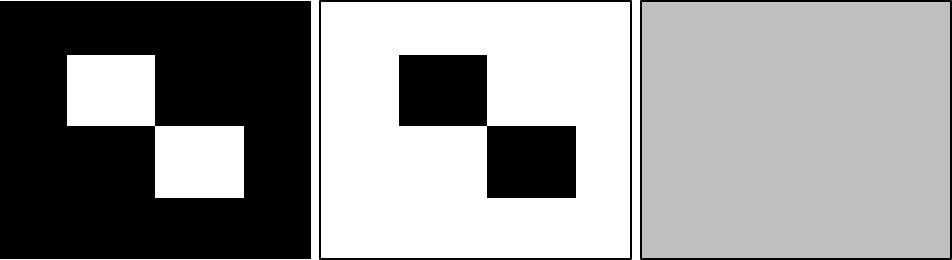}
	\caption{Illustration of an unregularized least square problem. The Euclidean distance between the third and first images are smaller than the distance between the first two images from the left. However, the third image (the plain gray image) does not contain any information regarding the first image.} 
	\label{fig:33}
\end{figure}

Note that the TV loss and the domain transfer loss both act as prior information to regularize the solutions of the least square loss terms. Least square minimization for an underdetermined problem is ill-posed and oftentimes does not converge to the desired solution. Hence, it is crucial to regularize the least square cost function. The toy example in Figure \ref{fig:33} illustrates the issue with an unregularized least square minimization. Also, it is important to train the Decoding and the Transfer Networks simultaneously. This ensures the regularization terms are imposed on both images produced by the Decoder and the Generator.
The coefficients $\lambda, \alpha, \beta$, and $\gamma$ are parameters of the model which should be tuned in order to obtain the desired balance among the incorporated loss terms. Next section describes the details of our implementation.
\section{Empirical Evaluation}
The training phase in our model consists of two steps. In the first step, we train the deep steganography algorithm in \cite{Baluja} using samples of ImageNet. In the second step, we train our steganalysis model using two sets of images. The first set is a small-sized dataset containing (secret, cover, container) image tuples. The container images were created by feeding the secret and cover images into a trained deep steganography algorithm. The second set is a larger dataset containing random sample images from ImageNet which do not necessarily overlap with the previously used training dataset. The second dataset is specifically used to train the Adversarial Network (the Discriminator). It is important to use a diverse dataset for the second set to avoid overfitting to any specific dataset. 

We used Adam optimization \cite{adam} to train all the networks in our model with parameters $\beta_1=0.9$ and $\beta_2=0.999$. Note that the Generator (the Transfer Network) needs to produce sample images that not only seem to be drawn from the space of real natural images, but are also close to the original secret images in Euclidean space. The additional cost terms, i.e., the conditional terms, narrow the space of feasible solutions which in turn makes it easier for the Discriminator (the Adversarial Network) to win the min-max game early in training. To balance the power between the two players, we trained the Generator multiple times (12 times) after each round of updating the Discriminator's weights. Additionally, we added noise to the Discriminator's target domain dataset (the second training set) which is the set of real natural images sampled from ImageNet. We gradually decrease the noise variance as the Generator catches up with learning. Similarly, the number of training iterations at each round of updating the Generator was gradually decreased as the Generator starts learning the steganalysis procedure. Note that once training phase is completed, the discriminator is detached from the model and it is not used for the evaluation phase.

To evaluate the steganalysis performance of our model, we tested our trained model on two different datasets. The first dataset was created by taking $30$ images with a smart phone. The second dataset contains $60$ samples of RGB images randomly drawn from ImageNet Bird, Sport, and Flowers categories. The test samples were not present in the training dataset used to train our model. In both cases, we down-sampled the images to $128\times 128$ resolution. We created container images by feeding the test dataset to the deep steganography algorithm and then provided our steganalysis model with the resulting container images as well as the original cover images. We compared the recovered images by our model with the original secret image to evaluate the steganalysis performance. Our evaluation was performed in both qualitative and quantitative forms.  Specifically, we used peak signal to noise ratio (PSNR) and the structural similarity index (SSIM) for the final recovered images. Also, we computed SSIM for the images obtained at the Decoder's output (The Transfer Network's input) to determine how much the domain adaption regularization has improved the quality of the decoded images.
\subsection{Simulation Results}
Our numerical results are summarized in Table \ref{table:1} which are obtained for the two test datasets. Our results suggest that the Transfer Network produces high quality colored revealed secret images with high SSIM. Notice that the improvement in SSIM for images produced by the Transfer Network over the images produced by the Decoder Network is non-trivial which highlights the effectiveness of our domain adaptation regularization technique. It is worth mentioning that our test results imply that the generator learns to not ignore the noise input which ultimately leads to a better generalization to different test datasets. For instance, in working with ImageNet dataset, we noticed that though the final transferred images were sharp images that were visually very similar to the original secret images, they featured slight variations in colors in the background of the Transferred images which did not exist the original secret images or in the Transfer Network's input images. Also, we noticed that stopping early in training helps the test performance, i.e., accepting some error in training phase results in better generalization to diverse test datasets.

\begin{figure*}[!ht]
	\centering
	\begin{subfigure}[]{} 
		\includegraphics[width=0.44\textwidth]{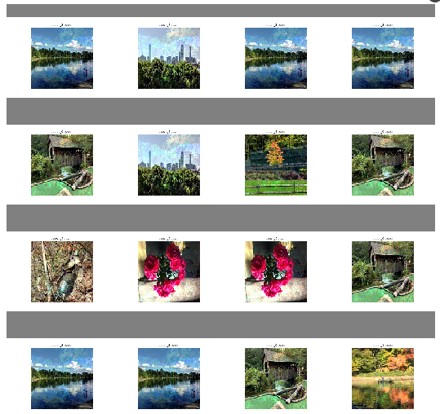}
	\end{subfigure} \hfill%
	\begin{subfigure}[]{} 
		\includegraphics[width=0.44\textwidth]{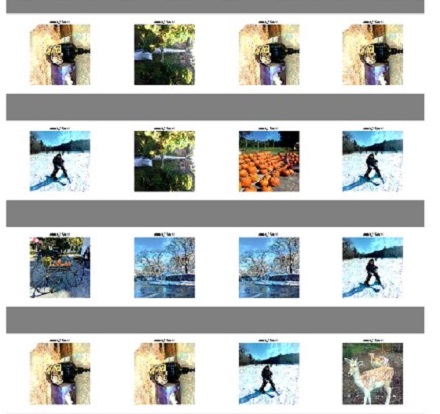}
	\end{subfigure}
	
	\vspace{1em} 
	\begin{subfigure}[]{} 
		\includegraphics[width=0.44\textwidth]{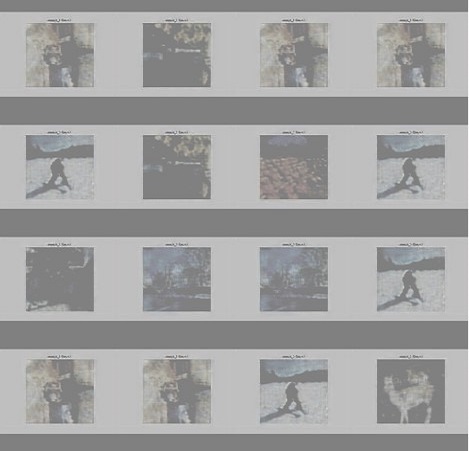}
\label{subfigure:3}		
	\end{subfigure}\hfill%
	\begin{subfigure}[]{} 
		\includegraphics[width=0.44\textwidth]{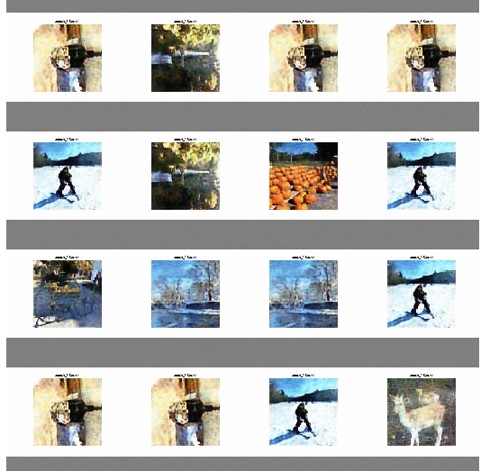}
\label{subfigure:4}		
		
	\end{subfigure}
	\caption{Our steganalysis results on sample images from a test dataset. Figure a and b show random samples of cover and secret images, respectively, drawn from our created dataset. Figure c displays the output of the Decoding Network in our model and Figure d displays the output of our Transfer Notwork, i.e., the final recovered secret images.} 
\label{figure:1}	
\end{figure*}

Figure \ref{figure:1} displays samples of original secret and cover images from our created dataset in the top row and the corresponding output images produced by our model. Specifically, Figure \ref{figure:1} (c) displays the decoded images produced by our Decoder network and and Figure \ref{figure:1} (d) displays the recovered secret images generated by our Transfer Network. Note that the mean square error cost term is applied into both the output of the Decoder Network and the Transfer Network. However, while the output of the Decoder Network features poor quality images with limited dynamic range and colors, the output of the Transfer Network features high quality images with rich colors and details visible to human eyes. The Transfer Network has learned the low level features of natural images such as colors and resolution and adjusted the decoded images accordingly which helped to recover the details of the embedded secret images.
\begin{figure*}[ht]
	\centering
		\includegraphics[width=0.7\textwidth]{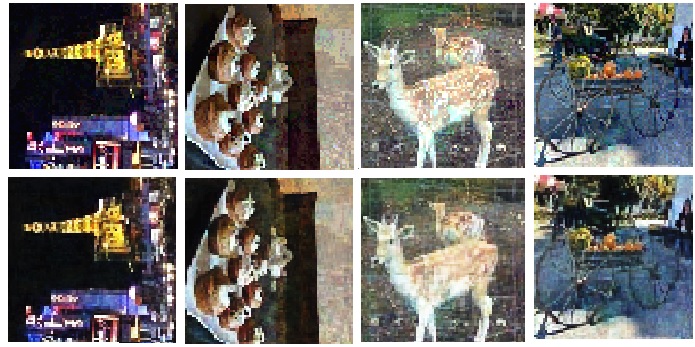}
	\caption{Sample recovered secret images from a test dataset. Top row shows samples of original secret images drawn from our created dataset and bottom row shows the corresponding final recovered secret images using our Decode and Transfer steganalysis model.} 
\label{figure:2}	
\end{figure*}
Additional samples of recovered secret images by our model are displayed in Figure \ref{figure:2} along with the original secret images. 
\begin{table}[t]
\begin{center}
\caption{Quantitative results on the performance of our steganalysis algorithm. The PSNR and SSIM quantities are calculated by comparison of the obtained images with corresponding original secret images. A Transferred image is the final recovered secret image which is an output of the Generator Network and a Decoded image is the middle output at the output of our Decoding Network.}
\label{table:1}
\vspace{0.1in}
\begin{tabular}{ | p {1.6cm}| p {1.6cm}| p{1.6cm} | p{1.6cm}|}
 \hline
 Test Dataset & Mean PSNR & Mean SSIM (Transferred) & Mean SSIM  (Decoded)\\
 \hline
 Our dataset  & 25.22 dB  & 0.89  & 0.59\\
\hline
Sampled ImageNet dataset   & 26.44 dB  & 0.82 & 0.65\\
\hline
\end{tabular}
\end{center}
\end{table}
\section{Discussion and Future Work}
In this work, we introduced a new steganalysis algorithm which targets deep embedded based steganography algorithms. We analyzed a specific steganography algorithm in which a full-sized secret image was embedded in the container image of the same size. We showed that not only our algorithm detects the embedded secret but it also fully recovers the secret image. Our empirical evaluation showed that with access to only a limited set of (secret, cover, container) images produced by the target steganography encoder, complete recovery of arbitrary secret images embedded by the same encoder is achievable.

Note that our model does not make any assumption on the steganography encoder, hence, our steganalysis algorithm can be successfully applied to attack an arbitrary deep embedded based steganography algorithm in which spatial information of the secret image is preserved and no cryptographic key was used. We propose our algorithm to be adopted as an \emph{attack model} in evaluation of future deep steganography algorithms. A potential approach to strengthen the secrecy of a steganography algorithm is to take into account our attack model while training the encoder. For instance, using an adversarial training scheme in which the encoder plays against our steganalysis model.

A key concept exploited in our steganalysis algorithm is simultaneously training the Decoding and the Transfer Networks. While with a limited training dataset, the Decoding Network is capable of partially decoding the secret images, it cannot produce high quality decoded images with all details visible to human eyes. In our model, we incorporated a domain adaptation regularization technique based on GAN to learn the prior information and solve the decoding problem jointly with the GAN's min-max regularization. The assumption here is that we have access to samples of the distribution from which secret images were drawn to train the Adversarial Network. Note that to train the Adversarial Network, we do not use the secret images used in training the target steganography algorithm. However, we assume different samples of the underlying distribution are known, e.g., samples of natural images  which is a reasonable assumption in most cryptoanalysis algorithms. The Adversarial Network learns the pattern of real images drawn from the secret distribution and through the zero-sum game helps the Transfer Network (the generator) to transfer poor quality decoded images to high quality RGB images with rich dynamic range and vivid details. The domain adaptation technique has been commonly used in style transfer application. In \cite{pixelda}, the domain adaptation was used to transfer binary images of hand written digits to colorful images of the digits. We introduced a new application of domain adaptation technique to improve the decoding process with application to steganalysis. 

One possible direction to extend this work is to apply our decoding and transfer techniques into a communication problem in which encoders and decoders are trained using only partial measurements (observations) from the communication channel. It is possible to adjust the Decoding Network to process the measurements and decode an estimate of the transmitted image which then transferred to high quality image by our Transfer Network. The side information that is provided by training the Transfer network can be beneficial in compression problems by reducing the bit-rate that is required to successfully transmit and decode a high quality RGB image (video). This model is essentially an example of Wyner-Ziv problem in which the decoder has access to the side information. Wyner-Ziv Theorem states that the input message can be effectively decoded by transmitting with a bit rate as low as the conditional rate of input message given the decoder's side information and this holds whether or not the transmitter has access to this side information \cite{wynerziv}. It would be interesting to quantify the bandwidth reduction achieved by incorporating our decode and transfer model in an image (video) communication problem, with or without security constraints.
\bibliography{example_paper}

\begin{thebibliography}{17}
\providecommand{\natexlab}[1]{#1}
\providecommand{\url}[1]{\texttt{#1}}
\expandafter\ifx\csname urlstyle\endcsname\relax
  \providecommand{\doi}[1]{doi: #1}\else
  \providecommand{\doi}{doi: \begingroup \urlstyle{rm}\Url}\fi

\bibitem[Aharon et~al.(2006)Aharon, Elad, and Bruckstein]{ksvd}
Aharon, M., Elad, M., and Bruckstein, A.
\newblock K-svd: An algorithm for designing overcomplete dictionaries for
  sparse representation.
\newblock \emph{IEEE Transactions on signal processing}, 54\penalty0
  (11):\penalty0 4311, 2006.

\bibitem[Baluja(2017)]{Baluja}
Baluja, S.
\newblock Hiding images in plain sight: Deep steganography.
\newblock In \emph{Advances in Neural Information Processing Systems}, pp.\
  2069--2079, 2017.

\bibitem[Bousmalis et~al.(2017)Bousmalis, Silberman, Dohan, Erhan, and
  Krishnan]{pixelda}
Bousmalis, K., Silberman, N., Dohan, D., Erhan, D., and Krishnan, D.
\newblock Unsupervised pixel-level domain adaptation with generative
  adversarial networks.
\newblock In \emph{The IEEE Conference on Computer Vision and Pattern
  Recognition (CVPR)}, pp.\ ~7, 2017.

\bibitem[Fridrich et~al.(2001)Fridrich, Goljan, and Du]{Frid}
Fridrich, J., Goljan, M., and Du, R.
\newblock Detecting lsb steganography in color, and gray-scale images.
\newblock \emph{IEEE multimedia}, 8\penalty0 (4):\penalty0 22--28, 2001.

\bibitem[Goljan et~al.(2014)Goljan, Fridrich, and Cogranne]{goljan}
Goljan, M., Fridrich, J., and Cogranne, R.
\newblock Rich model for steganalysis of color images.
\newblock In \emph{Information Forensics and Security (WIFS), 2014 IEEE
  International Workshop on}, pp.\  185--190. IEEE, 2014.

\bibitem[Hu et~al.(2018)Hu, Wang, Jiang, Zheng, and Li]{dcgan}
Hu, D., Wang, L., Jiang, W., Zheng, S., and Li, B.
\newblock A novel image steganography method via deep convolutional generative
  adversarial networks.
\newblock \emph{IEEE Access}, 6:\penalty0 38303--38314, 2018.

\bibitem[Isola et~al.(2017)Isola, Zhu, Zhou, and Efros]{styletransfer}
Isola, P., Zhu, J., Zhou, T., and Efros, A.~A.
\newblock Image-to-image translation with conditional adversarial networks.
\newblock In \emph{2017 IEEE Conference on Computer Vision and Pattern
  Recognition (CVPR)}, pp.\  5967--5976, 2017.

\bibitem[Kingma \& Ba(2015)Kingma and Ba]{adam}
Kingma, D.~P. and Ba, J.~L.
\newblock Adam: Amethod for stochastic optimization.
\newblock In \emph{International Conference on Learning Representations}, 2015.

\bibitem[Mielikainen(2006)]{MilkLSB}
Mielikainen, J.
\newblock Lsb matching revisited.
\newblock \emph{IEEE signal processing letters}, 13\penalty0 (5):\penalty0
  285--287, 2006.

\bibitem[Pevn{\`y} \& Fridrich(2007)Pevn{\`y} and Fridrich]{penvy2}
Pevn{\`y}, T. and Fridrich, J.
\newblock Merging markov and dct features for multi-class jpeg steganalysis.
\newblock In for Optics, I.~S. and Photonics (eds.), \emph{Security,
  Steganography, and Watermarking of Multimedia Contents IX}, pp.\  650503,
  2007.

\bibitem[Pevn{\`y} et~al.(2010)Pevn{\`y}, Filler, and Bas]{penvy}
Pevn{\`y}, T., Filler, T., and Bas, P.
\newblock Using high-dimensional image models to perform highly undetectable
  steganography.
\newblock In \emph{International Workshop on Information Hiding}, pp.\
  161--177. Springer, 2010.

\bibitem[Qian et~al.(2015)Qian, Dong, Wang, and Tan]{qian}
Qian, Y., Dong, J., Wang, W., and Tan, T.
\newblock Deep learning for steganalysis via convolutional neural networks.
\newblock In \emph{Media Watermarking, Security, and Forensics}, pp.\  94090J.
  International Society for Optics and Photonics, 2015.

\bibitem[T.~Bandyopadhyay et~al.(2013)T.~Bandyopadhyay, B, and
  Chatterji]{oracle}
T.~Bandyopadhyay, T., B, B.~B., and Chatterji, B.
\newblock Attacks on digital watermarked images in the internet environment and
  their counter measures.
\newblock \emph{International Journal of Advanced Research in Computer
  Science}, 4\penalty0 (2), 2013.

\bibitem[Tibshirani(1996)]{lasso}
Tibshirani, R.
\newblock Regression shrinkage and selection via the lasso.
\newblock \emph{Journal of the Royal Statistical Society. Series B
  (Methodological)}, pp.\  267--288, 1996.

\bibitem[Tikhonov et~al.(1995)Tikhonov, Goncharsky, Stepanov, and Yagola]{tikh}
Tikhonov, A., Goncharsky, A., Stepanov, V., and Yagola, A.~G.
\newblock Numerical methods for the solution of ill-posed problems (mathematics
  and its applications), 1995.

\bibitem[Wyner \& Ziv(1976)Wyner and Ziv]{wynerziv}
Wyner, A. and Ziv, J.
\newblock The rate-distortion function for source coding with side information
  at the decoder.
\newblock \emph{IEEE Transactions on information Theory}, 22\penalty0
  (1):\penalty0 1--10, 1976.

\bibitem[Zeng et~al.(2018)Zeng, Tan, Li, and Huang]{zeng}
Zeng, J., Tan, S., Li, B., and Huang, J.
\newblock Large-scale jpeg image steganalysis using hybrid deep-learning
  framework.
\newblock \emph{IEEE multimedia}, 13\penalty0 (5):\penalty0 1200--1214, 2018.

\end{thebibliography}
\bibliographystyle{icml2019}


\end{document}